\documentclass[runningheads]{llncs}

\usepackage[T1]{fontenc}
\usepackage{newtxtext,newtxmath}
\usepackage{microtype}
\usepackage{booktabs}
\usepackage{tabularx}
\usepackage{array}
\usepackage{xcolor}
\usepackage{listings}
\usepackage{tikz}
\usetikzlibrary{arrows.meta,positioning,fit,shapes.geometric}
\usepackage{url}
\usepackage{graphicx}
\usepackage{float}

\definecolor{codebg}{RGB}{247,247,247}
\definecolor{codeframe}{RGB}{215,215,215}
\definecolor{codekw}{RGB}{145,55,135}
\definecolor{codestr}{RGB}{175,75,75}
\definecolor{codecom}{RGB}{105,120,105}

\lstdefinestyle{swift}{
  language=Swift,
  basicstyle=\ttfamily\scriptsize,
  keywordstyle=\color{codekw},
  stringstyle=\color{codestr},
  commentstyle=\color{codecom},
  backgroundcolor=\color{codebg},
  frame=single,
  rulecolor=\color{codeframe},
  framerule=0.4pt,
  breaklines=true,
  showstringspaces=false,
  columns=fullflexible,
  keepspaces=true,
  xleftmargin=2pt,
  xrightmargin=2pt,
  aboveskip=5pt,
  belowskip=5pt
}
\newcommand{\aidcode}[1]{\texttt{#1}}

\begin{document}

\title{Reaching the Cards Apple Wallet Leaves Behind:\\Direct NFC Acquisition of PRO100, HUMO and UZCARD Payment Cards on iOS}
\titlerunning{Direct NFC Acquisition of Regional Payment Cards on iOS}

\author{Gusein Djalilov}
\authorrunning{G. Djalilov}
\institute{Independent Researcher, Tashkent, Uzbekistan\\
\email{g.djalilov@gmail.com}\\
ORCID: 0009-0007-3089-0867}

\maketitle

\begin{abstract}
Contactless payment from a phone has become routine in many markets, but the convenience is unevenly distributed. Apple currently lists Kazakhstan among supported Apple Pay markets, but not Uzbekistan, Kyrgyzstan, Tajikistan or Turkmenistan. In Uzbekistan, HUMO and UZCARD are the two national interbank retail card systems, and both include contactless card products. This paper describes an application-level card-capture method for tested PRO100, HUMO and UZCARD cards on NFC-capable iPhones. Using Core NFC, the reader opens a tag session, selects an application when necessary, and recovers the primary account number (PAN) and expiry date from the returned data. The difficult parts were empirical: identifying AIDs that worked on the tested cards and decoding card-generation-specific response layouts. The implementation was later hardened around a hybrid parsing path that prefers BER-TLV/EMV fields when present and keeps the observed offsets only as a legacy fallback. The paper also gives a system-level workflow, a threat model, explicit failure handling, a platform comparison, and a reproducible evaluation protocol. No claim is made that the same behavior is available unchanged on every iPhone/iOS combination, on iPadOS or macOS, or under future Core NFC policy. The method captures registration data; it does not emulate a card, authorize a payment, or treat an NFC read as proof of ownership.
\end{abstract}

\section{Introduction}

\subsection{The coverage gap}
Tap-to-pay on a smartphone is now ordinary in many markets, but availability is not uniform. At the time of writing, Apple's support list includes Kazakhstan but does not list Uzbekistan, Kyrgyzstan, Tajikistan or Turkmenistan as Apple Pay markets \cite{applepay}. The distinction matters because the work described here is not Apple Pay provisioning. It is card-data capture inside an application on an NFC-capable iPhone.

In Uzbekistan, the Central Bank identifies UZCARD and HUMO as the national retail bank-card systems \cite{cbu}. UZCARD reports more than 24.7 million cards in its network, while HUMO documents dual-interface contact/contactless cards and a large contactless acceptance infrastructure \cite{uzcardhome,humo,uzcardnfc}. PRO100 is a legacy Russian payment system rather than a dominant current Central Asian scheme: Sberbank stopped issuing PRO100 cards in 2016 and announced removal of its remaining PRO100 cards from circulation in 2024 \cite{interfax}. It is retained here only as a legacy technical case from the original test set.

\subsection{Why this matters}
The practical problem is small at the level of one user and large at the level of a card network. In an application that otherwise requires manual card registration, a user must copy account digits from a physical card. That introduces typing errors, extra steps and, in some implementations, accidental exposure through the ordinary text-input or clipboard path. Direct NFC capture can remove the typing step when the platform, device, entitlement configuration and card allow the tag to be read.

That does not make NFC capture inherently secure. The application still handles payment-account data and must minimize retention, avoid leaking values to logs or analytics, protect data while it moves through the app, and apply the same server-side verification used for manually entered details. The contribution is therefore intentionally narrow: an alternative input primitive for the tested environment.

\subsection{Contribution and scope}
This paper documents the engineering path rather than presenting the result as a new payment protocol. Concretely, it contributes: (i) a Core NFC tag-reader flow validated against the author's test set; (ii) empirical AID observations for HUMO and UZCARD cards, kept separate from publicly listed identifiers; (iii) a direct \textsc{READ RECORD} attempt followed, when necessary, by an application-selection fallback; (iv) recovery of PAN and expiry from observed response layouts; (v) a safer parser design that prefers known TLV/EMV fields before falling back to legacy offsets; and (vi) a threat model, failure model and reproducibility protocol that make the boundaries of the technique explicit.

Only the PAN and expiry required by the registration flow are interpreted. The method does not emulate a card, generate a cryptogram, perform a transaction, or replace issuer-side verification. The empirical AIDs and offset rules should be read as observations from a bounded test set, not as authoritative scheme specifications.

\section{Background}

\subsection{Core NFC and tag reader sessions}
Apple exposes supported contactless tag reading to third-party applications through Core NFC. The implementation uses \texttt{NFCTagReaderSession}, which can detect ISO 7816 and other supported tag types when the required entitlement and usage description are present \cite{corenfc}. ISO/IEC 14443-4 specifies the transmission protocol for contactless proximity objects, while ISO/IEC 7816-4 defines command and response structures, including APDUs \cite{iso14443,iso7816}.

There is an important platform boundary. Apple's current documentation states that \texttt{NFCTagReaderSession} does not support selection of payment-related application IDs; Apple points eligible payment-tag use cases in the European Union to \texttt{NFCPaymentTagReaderSession} \cite{corenfc}. The implementation described in this paper is therefore an empirical report from a tested environment, not a claim of unrestricted payment-card access on current or future iOS releases.

\subsection{APDU exchanges}
An APDU command contains a command header and, depending on the APDU case, optional command data and an expected-response length. A response may contain data followed by status bytes SW1 and SW2. These structures are defined by ISO/IEC 7816-4 \cite{iso7816}. The tested path uses only a small subset: \textsc{SELECT} by AID and \textsc{READ RECORD}.

A useful engineering distinction is that transport success is not the same thing as a usable application response. A production reader should check the Core NFC error, the status words, response length, and parser validity before accepting data. The early prototype relied too heavily on payload length; the revised design treats length as only one guard among several.

\subsection{Application identifiers and the regional schemes}
A smart card can expose more than one application, each selected by an application identifier (AID). Table~\ref{tab:aids} deliberately separates identifiers observed empirically on the author's HUMO and UZCARD test cards from a public PRO100 identifier. The empirical assignments should not be read as registry entries.

\begin{table}[H]
\caption{AIDs used or cross-checked in this study.}
\label{tab:aids}
\centering
\small
\begin{tabular}{lll}
\toprule
AID & Scheme & Basis / role \\
\midrule
\aidcode{A0860001000001} & HUMO & empirical; test-set primary \\
\aidcode{A0860001000002} & UZCARD & empirical; test-set primary \\
\aidcode{A0000004320001} & PRO100 & public-list identifier \\
\bottomrule
\end{tabular}
\end{table}

Two identifiers used in an earlier draft were mislabelled and are intentionally not used here. A public payment-industry AID catalogue identifies \aidcode{A0000006581010} as MIR Credit and \aidcode{A000000003000000} as a Visa Card Manager / GlobalPlatform-related identifier; the standard Visa debit/credit application is listed separately as \aidcode{A0000000031010} \cite{eftlab}. Treating the removed values as a HUMO/PRO100 fallback or as the Visa payment application would therefore be misleading.

\section{The Provisioning Gap in Detail}
Before turning to the method, it is useful to state why the ordinary routes do not solve this particular problem.

First, Apple Pay provisioning is a separate issuer-controlled path. Apple's \texttt{PKAddPaymentPassViewController} lets eligible apps add payment cards to Apple Pay, but the provisioning path requires Apple-granted capabilities and an issuer integration \cite{passkit}. Because Apple Pay is not currently listed as available in Uzbekistan, Kyrgyzstan, Tajikistan or Turkmenistan \cite{applepay}, this paper does not depend on Wallet provisioning for the domestic cards under study.

Second, EMV contactless specifications define standardized architecture, entry-point behavior and kernel processing for interoperable payment acceptance \cite{emv}. The response layouts encountered in the original test set did not match the assumptions of the parser then used in development. The implementation therefore treated the observed offsets as empirical. This point matters: it does not imply that the cards themselves are non-EMV; it means that the application's original extraction path was narrower than the range of data layouts the cards returned.

The application-level objective is simpler than payment acceptance. It is to capture the PAN and expiry fields that a user would otherwise type and pass them into an existing registration and server-verification flow. The NFC result is not a payment credential by itself and is not treated as evidence that the person holding the phone owns the card.

\section{Method}

\subsection{Session configuration}
The reader first checks that tag reading is available and then opens a session. Core NFC requires the appropriate tag-reader entitlement, a non-empty \texttt{NFCReaderUsageDescription}, and the relevant ISO 7816 application identifiers in the app configuration \cite{corenfc}. The listing uses generic names and is not tied to a particular commercial application.

\begin{lstlisting}
func beginSession() {
    guard NFCTagReaderSession.readingAvailable else {
        state = .cancelled
        lastError = "This device does not support tag reading."
        return
    }

    readerSession = NFCTagReaderSession(
        pollingOption: [.iso14443, .iso18092, .iso15693],
        delegate: self
    )
    readerSession?.alertMessage = "Hold your card near the phone."
    readerSession?.begin()
    state = .started
}
\end{lstlisting}

For this card flow, only ISO 14443/ISO 7816 processing is eventually accepted. The broader polling set reflects the original implementation and can be narrowed in a dedicated reader to reduce ambiguity.

\subsection{Tag detection and connection}
When a tag appears, the session connects to the first candidate and routes ISO 7816 tags to the card-processing path. Other tag types are rejected because they are outside the tested flow.

\begin{lstlisting}
func tagReaderSession(_ session: NFCTagReaderSession,
                      didDetect tags: [NFCTag]) {
    guard tags.count == 1, let tag = tags.first else {
        session.alertMessage = "Present one card at a time."
        session.restartPolling()
        return
    }

    session.connect(to: tag) { [weak self] error in
        guard let self else { return }
        if let error {
            session.invalidate(errorMessage: error.localizedDescription)
            self.state = .failed
            return
        }
        self.route(tag, session: session)
    }
}
\end{lstlisting}

The explicit one-tag check is a small but useful hardening step. It avoids guessing which object to process when several tags are visible.

\subsection{The two-command path}
The core method is a \textsc{READ RECORD} attempt plus a \textsc{SELECT}-by-AID fallback. The reader first attempts the record read. If the response is not usable, the reader selects a candidate application and then repeats the read. The listing below shows the empirically observed HUMO AID; the corresponding UZCARD or PRO100 candidate can be substituted when appropriate.

\begin{lstlisting}
let readRecordCmd = NFCISO7816APDU(
    data: Data([0x00, 0xB2, 0x01, 0x14, 0x00]))!

// HUMO candidate observed in the test set.
// Lc = 0x07, followed by seven AID bytes; final 0x00 is Le.
let selectAidCmd = NFCISO7816APDU(
    data: Data([0x00, 0xA4, 0x04, 0x00, 0x07,
                0xA0, 0x86, 0x00, 0x01,
                0x00, 0x00, 0x01, 0x00]))!
\end{lstlisting}

The original implementation used payload length as a pragmatic signal for entering the fallback path. That was convenient during development, but it is not a protocol-level proof that selection is required. The revised rule is stricter: transport errors and status words are checked first; malformed or implausibly short data is never passed to the card parser; and a fallback AID is tried only from a small allowlist appropriate to the expected scheme.

\subsection{Field recovery: TLV first, offsets second}
The original parser converted the response to hexadecimal text and recovered the account number and expiry by position. Two layouts appeared in the test set, so the parser switched offsets when the primary PAN region contained non-decimal hexadecimal characters.

\begin{lstlisting}
func extractLegacyCard(from hex: String) -> PaymentCard? {
    guard hex.count >= 60 else { return nil }
    let start = hex.startIndex

    func slice(_ a: Int, _ b: Int) -> String {
        let lo = hex.index(start, offsetBy: a)
        let hi = hex.index(start, offsetBy: b)
        return String(hex[lo..<hi])
    }

    var pan = slice(10, 26)
    var validity = slice(27, 31)

    if pan.rangeOfCharacter(from: .letters) != nil {
        pan = slice(44, 60)
        validity = slice(34, 38)
    }

    guard pan.allSatisfy(\.isNumber),
          let (year, month) = splitEvenly(validity) else { return nil }

    return PaymentCard(pan: pan, expMonth: month, expYear: year)
}
\end{lstlisting}

Those offsets remain useful for reproducing the behavior of the legacy test set, but they should not be the first parser a new implementation reaches for. A more resilient path is:

\begin{enumerate}
\item Parse the returned data as bounded BER-TLV where the structure permits it.
\item Look for known EMV fields such as tag \aidcode{5A} (PAN), \aidcode{5F24} (application expiration date), or \aidcode{57} (Track 2 equivalent data) and validate BCD content and field lengths \cite{emv}.
\item If the response is not a usable TLV structure, apply a scheme/card-generation-specific legacy decoder only when the response matches a previously observed signature.
\item Reject anything else and return the user to manual registration rather than guessing.
\end{enumerate}

This keeps the empirical contribution of the original implementation while reducing the chance that a new issuer profile silently shifts the targeted bytes. It also makes the fallback visible in code review instead of hiding it inside a generic string-slicing routine.

\subsection{End-to-end acquisition workflow}
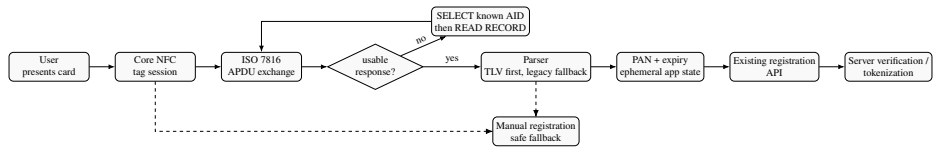
\begin{figure}[H]
\centering
\resizebox{\textwidth}{!}{%
\begin{tikzpicture}[
  node distance=8mm and 8mm,
  box/.style={draw, rounded corners, align=center, minimum height=9mm, minimum width=25mm, fill=gray!5},
  decision/.style={draw, diamond, aspect=2.1, align=center, inner sep=1.5pt, fill=gray!5},
  arrow/.style={-{Latex[length=2mm]}, thick},
  fallback/.style={-{Latex[length=2mm]}, dashed}
]
\node[box] (user) {User\\presents card};
\node[box, right=of user] (session) {Core NFC\\tag session};
\node[box, right=of session] (apdu) {ISO 7816\\APDU exchange};
\node[decision, right=of apdu] (valid) {usable\\response?};
\node[box, above right=6mm and 10mm of valid] (select) {SELECT known AID\\then READ RECORD};
\node[box, right=18mm of valid] (parse) {Parser\\TLV first, legacy fallback};
\node[box, right=of parse] (fields) {PAN + expiry\\ephemeral app state};
\node[box, right=of fields] (api) {Existing registration\\API};
\node[box, right=of api] (server) {Server verification /\\tokenization};
\node[box, below=11mm of parse] (manual) {Manual registration\\safe fallback};

\draw[arrow] (user) -- (session);
\draw[arrow] (session) -- (apdu);
\draw[arrow] (apdu) -- (valid);
\draw[arrow] (valid) -- node[above]{yes} (parse);
\draw[arrow] (valid) -- node[sloped,above]{no} (select);
\draw[arrow] (select.west) -| (apdu.north);
\draw[arrow] (parse) -- (fields);
\draw[arrow] (fields) -- (api);
\draw[arrow] (api) -- (server);
\draw[fallback] (session.south) |- (manual.west);
\draw[fallback] (parse.south) -- (manual.north);
\end{tikzpicture}}
\caption{High-level NFC acquisition and card-registration workflow. Failure at the NFC or parsing stage falls back to the existing manual path.}
\label{fig:arch}
\end{figure}

Figure~\ref{fig:arch} shows the complete path. The important boundary is after parsing: NFC capture feeds the same registration and server-verification flow used by manual entry. It does not create a separate trust path.

\section{Security Model}

\subsection{Assets and trust boundaries}
The primary asset is the PAN. The expiry date is also payment-account data when associated with the PAN, and the raw APDU payload can contain additional data that the application has no reason to keep. The trust boundaries are: (i) physical card to NFC controller/Core NFC; (ii) Core NFC callback data to the application process; and (iii) application process to the registration backend.

PCI DSS is relevant because it is intended for entities that store, process or transmit cardholder data, or that can affect the security of that data \cite{pcidss}. Whether a particular mobile architecture is in PCI scope depends on the wider system and cannot be decided from the code fragment alone. The safe design choice is nevertheless the same: minimize the amount of account data the mobile application handles and minimize how long it handles it.

\subsection{Threats and mitigations}
Table~\ref{tab:threats} summarizes the main threats considered by the revised design.

\begin{table}[H]
\caption{Threat model for NFC-assisted registration.}
\label{tab:threats}
\centering
\scriptsize
\begin{tabularx}{\textwidth}{>{\raggedright\arraybackslash}p{2.35cm} >{\raggedright\arraybackslash}X >{\raggedright\arraybackslash}X}
\toprule
Threat & Failure mode & Mitigation in the design \\
\midrule
Replay of captured data & A previously captured PAN/expiry pair is resubmitted and incorrectly treated as proof that the card is present. & NFC capture is only an input method. The backend applies the same ownership/registration verification as manual entry; no trust decision is based on the read alone. \\
\addlinespace
Malicious or unrelated tag & Arbitrary tag data reaches a parser that assumes a payment-card layout. & Accept only the expected tag technology, allowlist candidate AIDs, validate status words and lengths, use bounded TLV parsing, and reject unknown layouts. \\
\addlinespace
Parser confusion & A new card generation shifts fields and a fixed offset extracts the wrong bytes. & Prefer BER-TLV/EMV tags; permit offset parsing only for a recognized legacy layout; validate PAN digits, length and expiry before use. \\
\addlinespace
Memory/log leakage & PAN is copied into debug logs, analytics events, crash metadata or long-lived strings. & Do not log APDU payloads or PAN, keep raw \texttt{Data} short-lived, avoid unnecessary string copies, and clear references as soon as registration has been handed off. \\
\addlinespace
Persistent local exposure & Card data is stored on device longer than necessary. & Do not persist raw APDU responses. Prefer not to persist PAN at all; store only server-issued identifiers/tokens where the product architecture allows it. \\
\addlinespace
Tag loss / interrupted read & A partial exchange is accepted as a complete result. & Treat transport errors, tag loss, unexpected status words and incomplete parse as failure; restart the session or fall back to manual entry. \\
\bottomrule
\end{tabularx}
\end{table}

One limitation is worth stating explicitly. In a high-level language and managed runtime, "zeroing memory" is not a blanket guarantee: copies can be introduced by strings, framework bridges or temporary values. The practical goal is to reduce copies and lifetime, keep sensitive values out of logs and persistent stores, and avoid retaining the full response once the two required fields have been validated.

\section{Failure Scenarios and Recovery}
The happy path is short; the failure paths are where a reader becomes production software. The revised implementation treats the following cases separately rather than collapsing them into a generic "NFC failed" message.

\begin{itemize}
\item \textbf{NFC unavailable or policy-restricted.} Hide or disable NFC capture and leave manual registration available. A user should not be trapped in a flow that depends on a capability the device or current OS policy does not expose.
\item \textbf{More than one tag detected.} Restart polling and ask for one card. Selecting the first tag is easy to code but difficult to justify.
\item \textbf{Connection error or tag loss.} Invalidate the current exchange and allow a fresh attempt. A partial response is never reused.
\item \textbf{Unexpected APDU status.} Do not parse the response body as card data. Where the implementation has a well-defined alternative candidate AID, it can try that candidate; otherwise it fails closed.
\item \textbf{Malformed TLV or unknown response layout.} Reject the payload. The legacy offset parser is entered only for a response signature that was already observed and tested.
\item \textbf{Field validation failure.} PAN and expiry must pass basic structural checks before they are sent to the registration backend. A parser result that cannot pass those checks is treated as no result.
\end{itemize}

This recovery strategy is intentionally conservative. Manual input already exists, so there is little value in making the NFC path "succeed" on ambiguous data.

\section{Practical Challenges}

\subsection{Identifier discovery}
The HUMO and UZCARD values in Table~\ref{tab:aids} were assembled empirically by trying candidate identifiers against the author's test cards and retaining values that produced the expected application response. They are measurements, not registry assignments. The public AID list used for cross-checking gives \aidcode{A0000004320001} for PRO100 and assigns two values used incorrectly in an earlier draft to other applications \cite{eftlab}. Keeping observed behavior separate from documented identity is essential for reproducibility.

\subsection{Decoding}
Once a tested card responds, the original parser did not extract the targeted fields through its normal EMV path. PAN and expiry were therefore located empirically by offset, and at least two response layouts appeared in the test set. That explains the original branch that switches offsets when the expected decimal content is absent.

The revised parser does not throw those observations away; they remain useful for the cards on which they were discovered. What changes is the order of trust. A structured TLV interpretation is attempted first, and offset decoding becomes a narrow compatibility path. This is a more maintainable response to future card generations than continuously adding new magic offsets to a single function.

\section{Experimental Evaluation and Reproducibility}

\subsection{What was actually validated}
On the author's tested iPhone/iOS configurations, the method recovered PAN and expiry from the tested PRO100, HUMO and UZCARD cards and supplied those fields to an application registration flow. This is the result the original development work can support.

The development work predated the paper and was not instrumented as a controlled laboratory study. Card-by-card research metadata, aggregate success counters and APDU timing distributions were not retained. For that reason this paper does not invent a sample size, a success percentage, a median latency, or a device/version coverage matrix after the fact. The claim is deliberately functional and bounded to the test set. A statistically meaningful compatibility claim would require a new measurement run.

\subsection{Protocol for a follow-up measurement run}
A reproducible evaluation should record more than whether one card worked once. Table~\ref{tab:metrics} defines the measurements intended for a follow-up run and makes the reporting criteria explicit.

\begin{table}[H]
\caption{Recommended evaluation measurements for a controlled replication.}
\label{tab:metrics}
\centering
\scriptsize
\begin{tabularx}{\textwidth}{>{\raggedright\arraybackslash}p{2.3cm} >{\raggedright\arraybackslash}X >{\raggedright\arraybackslash}X}
\toprule
Metric & Definition & Reporting rule \\
\midrule
Card coverage & Number of physical cards per scheme, issuer and card generation. & Report unique cards, not repeated reads of the same card. \\
\addlinespace
Success rate & Valid PAN+expiry captures divided by initiated read attempts. & Report per scheme and per device/OS combination, with attempt count. \\
\addlinespace
Acquisition latency & Time from tag detection to validated card fields. & Report median and at least p95; separate first read from AID-fallback reads. \\
\addlinespace
Retry rate & Attempts requiring reconnect, repoll or AID fallback. & Separate tag-loss retries from parser/protocol fallbacks. \\
\addlinespace
Device coverage & iPhone model and NFC capability. & Record hardware model for every run. \\
\addlinespace
iOS coverage & Exact iOS build/version. & Record version because Core NFC behavior and policy can change across releases. \\
\addlinespace
Failure category & Unsupported tag, connection error, status-word rejection, malformed TLV, unknown layout, field validation failure. & Count categories separately instead of one generic failure total. \\
\bottomrule
\end{tabularx}
\end{table}

The run should use repeated attempts per physical card and randomize card/device order where practical. Raw PAN values should not be included in the research dataset. A card can be represented by a study-local identifier and, if needed, a non-reversible test label assigned manually before the run.

\subsection{Comparison with other registration approaches}
Table~\ref{tab:comparison} positions the method against two common alternatives. The comparison is qualitative because no controlled timing study was preserved from the original development.

\begin{table}[H]
\caption{Qualitative comparison of card-registration approaches.}
\label{tab:comparison}
\centering
\scriptsize
\begin{tabularx}{\textwidth}{>{\raggedright\arraybackslash}p{2.35cm} >{\raggedright\arraybackslash}X >{\raggedright\arraybackslash}X >{\raggedright\arraybackslash}X}
\toprule
Approach & Strength & Main limitation & Trust implication \\
\midrule
Manual card entry & Works without NFC access and is broadly portable across devices. & Typing friction; possible input errors; data passes through the normal text-entry path. & Still requires server-side ownership/registration verification. \\
\addlinespace
iOS Core NFC capture & Removes keyboard entry and can acquire the required fields in one physical interaction on supported configurations. & Core NFC policy, entitlement requirements, device/OS behavior and card layout can block the path. & Read is an input convenience, not proof of ownership. \\
\addlinespace
Android ISO-DEP path & Android exposes \texttt{IsoDep.transceive()} for raw ISO-DEP request/response I/O, leaving the application to implement its protocol layer \cite{androidisodep}. & Scheme behavior, parser correctness and device/card compatibility still require validation; this paper did not benchmark an Android implementation. & Same server verification remains necessary. \\
\bottomrule
\end{tabularx}
\end{table}

The Android comparison is deliberately limited to the platform API model. It would be misleading to claim parity or superiority without implementing and measuring the same card set on both platforms.

\subsection{Sanitized APDU sequence}
For reproducibility, the command sequence can be published without publishing cardholder data. The sequence below is the one used by the prototype. Response bodies are intentionally omitted because the original development did not create a research dataset with safely anonymized payloads, and PAN-bearing responses should not be reconstructed for publication.

\begin{verbatim}
C-APDU  00 B2 01 14 00
R-APDU  <response body redacted>  <SW1> <SW2>

fallback when the first response is not usable:

C-APDU  00 A4 04 00 07 A0 86 00 01 00 00 01 00
R-APDU  <response body redacted>  <SW1> <SW2>
C-APDU  00 B2 01 14 00
R-APDU  <response body redacted>  <SW1> <SW2>
\end{verbatim}

A future supplementary dataset should log only what is needed for reproducibility: command bytes, status words, timing, parser path and failure category. If response bodies are retained for research, they need a separate redaction/tokenization process and a clear data-handling protocol rather than ad hoc masking in a paper draft.

\section{Results and Discussion}
Within the validated environment, the user-facing effect is to replace manual PAN/expiry entry with NFC capture. The application receives the APDU response but intentionally interprets only the fields needed for registration. Other returned bytes are not intentionally retained. The method neither emulates a card nor performs a transaction.

The hybrid parsing strategy also clarifies what is general and what is not. TLV-aware extraction is the preferred path when a recognizable structure is present. The observed offsets remain empirical compatibility rules for particular response layouts. If a new card generation matches neither path, the correct outcome is failure and manual entry, not speculative parsing.

The largest limitation remains platform policy. Apple's current documentation explicitly says that \texttt{NFCTagReaderSession} does not support selection of payment-related AIDs and points eligible EU payment-tag use cases to \texttt{NFCPaymentTagReaderSession} \cite{corenfc}. Device support, entitlements, OS policy and region can therefore determine whether the path is available regardless of the APDU logic itself. The technique should be understood as a tested engineering result under bounded conditions, not as a stable public contract guaranteed by iOS.

A second limitation is experimental. The original implementation proved feasibility but did not preserve the logs required for a quantitative study. The evaluation protocol in Section~8 is included so that a future iteration can make stronger claims with measured coverage, latency and error distributions instead of relying on retrospective estimates.

\section{Conclusion}
Apple's support list still omits Uzbekistan, Kyrgyzstan, Tajikistan and Turkmenistan from Apple Pay availability \cite{applepay}, while Uzbekistan's domestic card infrastructure operates at tens-of-millions-of-issued-cards scale \cite{uzcardhome}. Within the iPhone/iOS test environment described here, direct Core NFC capture reduced a manual registration step to an NFC interaction for the tested PRO100, HUMO and UZCARD cards.

The main engineering findings are empirical AID selection for the tested HUMO and UZCARD cards and response decoding that originally depended on observed offsets. The revised design keeps those findings but surrounds them with stricter protocol checks, a TLV-first parsing strategy, explicit failure recovery, and a security model that treats PAN capture as sensitive input handling rather than a trust signal. PRO100 remains a legacy test case.

The next useful step is not another parser heuristic. It is a controlled replication across a documented set of cards, iPhone models and iOS versions, with success rate, latency, retries and failure categories recorded from the start. That would turn a successful production engineering technique into a stronger experimental result without overstating what the original data can prove.

\section*{Acknowledgment}
The author thanks the reviewer of an earlier draft for pushing the work beyond the implementation details, particularly on experimental reporting, security boundaries, reproducibility and long-term parser resilience.

\end{document}